\documentclass[conference]{IEEEtran}
\usepackage{spconf,amsmath,graphicx}
\usepackage{booktabs}
\usepackage{spconf,amsmath,graphicx}
\usepackage{mathtools}
\usepackage{xcolor}
\usepackage{amsfonts}
\usepackage{xfrac}
\usepackage{amssymb}
\usepackage{array}
\usepackage{multirow}
\usepackage{varwidth}
\usepackage{algorithm,algorithmicx}
\usepackage{algpseudocode}
\usepackage{cite}
\usepackage{amsmath,amssymb,amsfonts}
\usepackage{graphicx}
\usepackage{textcomp}
\usepackage{xcolor}
\usepackage{enumitem}
\setlist{nosep, leftmargin=14pt}

\usepackage{mwe} 

\usepackage{spconf}
\usepackage{mwe} 

\DeclareMathOperator*{\argmin}{arg\,min}

\newcommand{\set}[1]{\mathcal{#1}}

\newcommand{\setX}{\set{X}}

\title{Towards Causality-Aware Modeling for Multimodal Brain-Muscle Interactions}

\name{Farwa Abbas$^{\dagger}$, Wei Dai$^{\dagger}$, Zoran Cvetkovic$^{\ast}$, and Verity McClelland$^{\ast}$ }
 \address{ $^{\dagger}$Imperial College London,
     $^{\ast}$King's College London\\
        \{f.abbas20, wei.dai1\}@imperial.ac.uk, \{zoran.cvetkovic,verity.mcclelland\}@kcl.ac.uk}
\begin{document}
%
\maketitle


\begin{abstract}
Robust characterization of dynamic causal interactions in multivariate biomedical signals is essential for advancing computational and algorithmic methods in biomedical imaging. Conventional approaches, such as Dynamic Bayesian Networks (DBNs), often assume linear or simple statistical dependencies, while manifold-based techniques like Convergent Cross Mapping (CCM) capture nonlinear, lagged interactions but lack probabilistic quantification and interventional modeling. We introduce a DBN-informed CCM framework that integrates geometric manifold reconstruction with probabilistic temporal modeling. Applied to multimodal EEG–EMG recordings from dystonic and neurotypical children, the method quantifies uncertainty, supports interventional simulation, and reveals distinct frequency-specific reorganization of cortico-muscular pathways in dystonia. Experimental results show marked improvements in predictive consistency and causal stability as compared to baseline approaches, demonstrating  the potential of causality-aware multimodal modeling for developing quantitative biomarkers and guiding targeted neuromodulatory interventions.
\end{abstract}

\section{Introduction}
Understanding causal relationships within complex biological and physiological systems is essential for accurate diagnosis, treatment planning, and prognostic evaluation. These systems when observed through functional Magnetic Resonance Imaging (fMRI), Electrophysiology (EEG), Electromyography (EMG), or other dynamic imaging modalities, exhibit intricate nonlinear dynamics, feedback loops, and stochastic fluctuations. Traditional linear or purely deterministic approaches prove insufficient for characterizing such systems, as they cannot fully account for hidden physiological states, temporal delays, and dynamically evolving interactions.

Correlation-based approaches \cite{kim2017cross, ullah2025skills, ahmad2025future, abbas2024robust, zhang2025regimefolio, abbas2025infr} can detect statistical associations but cannot distinguish direct causal influences from indirect effects or shared confounding inputs. Such limitations are especially consequential in biomedical contexts where interventional reasoning is required to inform therapeutic actions. Dynamic Bayesian Networks (DBNs) address this challenge by supporting probabilistic causal inference consistent with intervention semantics \cite{shiguihara2021dynamic}. They can also scale effectively to multivariate neuroimaging and physiological time-series data by exploiting structured sparsity, anatomical priors, and approximate inference to constrain the otherwise high-dimensional temporal dependency space. However, standard DBNs often impose linear-Gaussian assumptions that fail to represent nonlinear and delay-driven coupling mechanisms prevalent in neural and neuromuscular systems. As a result, they may under-detect genuine causal influences arising from oscillatory, partially deterministic, or manifold-constrained dynamics \cite{sanchez2022causal}. To overcome these limitations, approaches grounded in nonlinear dynamical systems theory model complex system behavior as trajectories on low-dimensional attractor manifolds, capturing intrinsic temporal and deterministic structure \cite{frassineti2022analysis}. Within this theoretical foundation, delay-embedding techniques reconstruct the underlying state-space geometry from time series measurements, making hidden deterministic structure and directional coupling observable. Leveraging these concepts, Convergent Cross Mapping (CCM) employs attractor reconstruction to identify nonlinear causal interactions by assessing the cross-predictability between state spaces. As a result, CCM is capable of revealing subtle directional couplings even when they manifest as weak or lagged influences within complex feedback processes \cite{avvaru2023effective}. While CCM is highly effective in detecting nonlinear and lag-dependent coupling in physiological systems, it remains fundamentally an observational method, lacking a joint generative framework, measures to quantify uncertainty, and the capacity for counterfactual or interventional reasoning critical to translational biology and medicine \cite{deng2024causalized, abbas2025scalar}. 

Previous works have made efforts to combine CCM with Gaussian processes \cite{butler2023causal} and information-theoretic measures \cite{mccracken2014convergent,ahmad2025resilient, chen20253s, goel2025co, jois2026australian, barraquand2021inferring} to provide confidence intervals or probabilistic estimates of influence, partially addressing the inability of CCM to quantify uncertainty. However, extending CCM to enable probabilistic reasoning and interventional causal analysis remains largely unexplored. Therefore, to address the shortcomings of purely probabilistic and deterministic methods while leveraging their strengths, we introduce a DBN-Informed CCM framework that combines the complementary strengths of manifold-based nonlinear causal discovery with probabilistic temporal modeling. In this framework, CCM-generated coupling strengths guide DBN structure learning, highlighting nonlinear, lagged, or manifold-constrained interactions often missed by standard probabilistic methods. Conversely, conditional probabilities derived from the DBN refine CCM predictions, allowing manifold-based inference to account for stochasticity and uncertainty in biomedical time series. This ensures that the geometric relationships on reconstructed attractor manifolds inform probabilistic structure discovery, while temporal dependencies captured by the DBN enhance the reliability of manifold-based causal inference. To avoid \textit{double-counting} causal evidence as a result of circularity, we implement careful normalization of CCM-derived coupling strengths before incorporating them as priors in DBN structure learning, ensuring that the influence of each source is properly weighted and does not reinforce itself through feedback. We employ a novel dual-weight learning scheme, in which kernel-based manifold similarity informs DBN structural priors while the temporal conditional probabilities capture stochastic dependencies. The key contributions of this work are as follows:

\begin{enumerate}
    \item We propose a unified causal modeling framework combining the detection of intricate causal relationships with uncertainty quantification and interventional capabilities, providing an effective framework for investigating multiscale, nonlinear dynamics in physiological systems.

    \item We propose a novel way to quantify causal effect magnitude via both probabilistic and geometric changes, supporting functional hypothesis testing from observational data.

    \item We experimentally validate our proposed method on EEG and EMG recordings from dystonic and healthy children. By analyzing cross-modal connectivity across frequency bands, our method uncovers interpretable neural-muscle causal pathways consistent with motor physiology.
\end{enumerate}

\section{Proposed Methodology}

Consider two time series $X = \{x_1, ..., x_N\}$ and $Y = \{y_1, ..., y_N\}$ representing observations from potentially coupled dynamical systems. Our approach leverages Takens' embedding theorem, which establishes that under suitable smoothness conditions and generic observables, the reconstructed manifold preserves the essential topological and dynamical properties of  attractor of the original system. Specifically, each point in the shadow manifold $M_X[i]$ is constructed through delay embedding:
\begin{equation}
    M_X[i] = [x_i, x_{i-\tau}, ..., x_{i-(E-1)\tau}]
\end{equation}
The fidelity of this reconstruction critically depends on two parameters: the embedding dimension $E$ and time delay $\tau$. Their selection follows from both theoretical requirements and practical considerations.
\begin{algorithm}[t]
\caption{DBN-Informed CCM}
\begin{algorithmic}
    \State \textbf{Input:} Time series $X = \{x_1, ..., x_N\}$, $Y = \{y_1, ..., y_N\}$, embedding dimension $E$,
    \State \quad time delay $\tau$, DBN structure $\mathcal{G}$
    
    \State $M_X[i] = [x_i, x_{i-\tau}, ..., x_{i-(E-1)\tau}]$ \Comment{Construct shadow manifold $M_X$ with each point $M_X[i]$}
    
    \For{$t \leftarrow 1$ to $N$}
        \State $\mathcal{I} = \argmin_{j \in J, |J|=E+1} \sqrt{\sum_{k=1}^E (M_X[t]_k - M_X[j]_k)^2}$ \Comment{Find $E+1$ nearest neighbor indices}
        
        \State $u_i = \exp(-\frac{d_{ti}^2}{2\sigma^2}), \quad i \in \mathcal{I}$ \Comment{Compute kernel weights for neighbors}
        
        \State $p_i = P(Y[i] | \text{Pa}(Y[i])), \quad i \in \mathcal{I}$ \Comment{Get DBN conditional probabilities}
        
        \State $w_i = \frac{u_i p_i}{\sum_{j \in \mathcal{I}} u_j p_j}, \quad i \in \mathcal{I}$ \Comment{Compute normalized combined weights}
        
        \State $Y_{pred}[t] = \sum_{i \in \mathcal{I}} w_i Y[i]$ \Comment{Predict Y value as weighted sum}
    \EndFor
    
    \State $\rho = \frac{\sum_{t=1}^N (Y[t] - \bar{Y})(Y_{pred}[t] - \bar{Y}_{pred})}{\sqrt{\sum_{t=1}^N (Y[t] - \bar{Y})^2} \sqrt{\sum_{t=1}^N (Y_{pred}[t] - \bar{Y}_{pred})^2}}$ \Comment{Compute correlation between $Y$ and $Y_{pred}$}
    
    \State \textbf{Output:} Correlation $\rho_{X \rightarrow Y}$
\end{algorithmic}
\label{algo8}
\end{algorithm}
In the reconstructed state space, we identify the optimal set of $E+1$ nearest neighbors for each point through:
\begin{equation}
    \mathcal{I} = \argmin_{j\in J,|J|=E+1} \sqrt{\sum_{k=1}^E (M_X[i]_k - M_X[j]_k)^2}
\end{equation}
The number of neighbors selected will be $E+1$ representing the minimal set required for unique point determination in $E$-dimensional space plus one additional point for robust estimation. Our hybrid weighting scheme integrates geometric proximity with probabilistic causality in a novel way through two fundamental components:
\begin{equation}
    u_i = \exp(-\frac{d_i^2}{2\sigma^2}), \quad
    p_i = P(Y[i]|\text{Pa}(Y[i])), \quad i \in \mathcal{I}
\end{equation}
The exponential kernel weights $u_i$ are employed due to their inherent
distance-decay characteristics, which align with manifold diffusion principles
and provide stable local density estimates in noisy environments. In parallel,
the DBN-derived conditional probabilities $p_i$ encode the learned temporal
causal structure, offering a probabilistic constraint on state transitions.
These components are integrated through normalized weights as:
\begin{equation}
    w_i = \frac{u_ip_i}{\sum_{i\in\mathcal{I}} u_ip_i}, \quad Y_{pred}[t] = \sum_{i \in \mathcal{I}} w_i Y[i]
\end{equation}
This normalization ensures proper probabilistic interpretation while preserving the relative influence of both geometric and causal factors. The prediction framework employs these weights in a locally linear approximation scheme.
The strength of causal relationships is quantified through correlation analysis:
\begin{equation}
    \rho = \frac{\sum_{i=1}^N (Y[i]-\bar{Y})(Y_{pred}[i]-\bar{Y}_{pred})}{\sqrt{\sum_{i=1}^N (Y[i]-\bar{Y})^2\sqrt{\sum_{i=1}^N (Y_{pred}[i]-\bar{Y}_{pred})^2}}}
\end{equation}
\begin{figure}[t]
    \centering
    \includegraphics[width=0.7\linewidth]{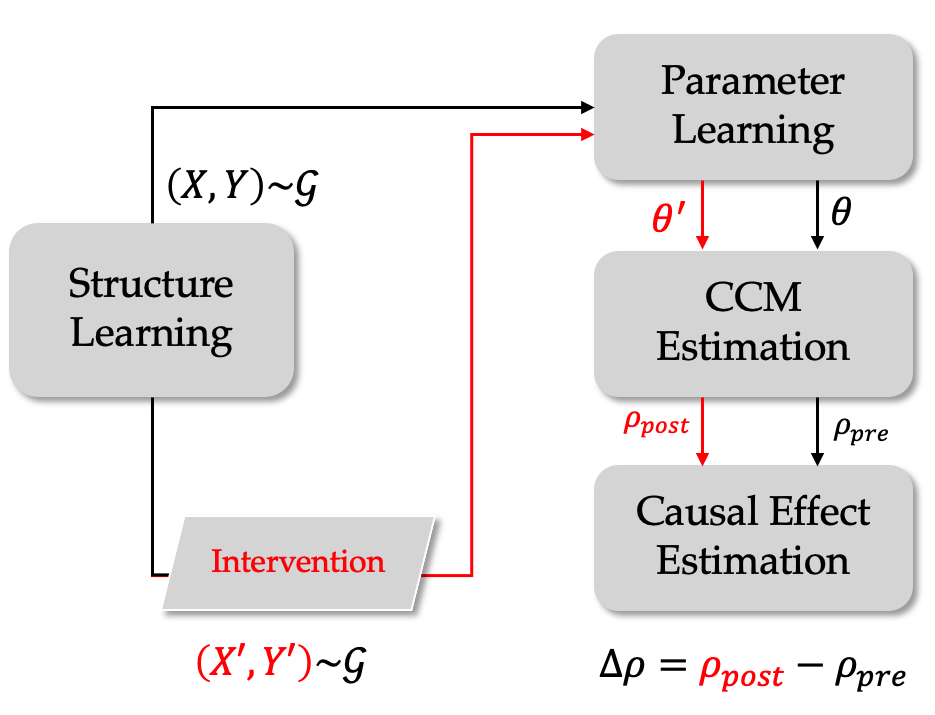}
    \caption{Schematic of proposed DBN-Informed CCM}
    \label{dbn}
    \vspace{-1\baselineskip}
\end{figure}
The complete algorithmic implementation is detailed in Algorithm \ref{algo8}. Interventions are designed to capture both local and global changes in system dynamics, enabling robust causal effect estimation across multiple temporal scales. This dual-scale approach proceeds along different dimensions.

First, geometric changes in the shadow manifold $M_X$ reflect local perturbations in the state-space structure of system. Post-intervention, points in $M_X[i]$ exhibit reorganized nearest-neighbor relationships, capturing how genuine causal influences manifest as measurable modifications in the manifold geometry. Second, we track changes in DBN conditional probabilities $P(Y[i]|\text{Pa}(Y[i]))$. Probabilistic alterations propagate through the hybrid weights $w_i$, revealing causal effects that influence prediction accuracy. The integration of geometric and probabilistic analyses provides a more complete characterization of causal effects than either method alone. Causal effect estimation is formalized by comparing pre- and post-intervention predictability:
\begin{align}
    \rho_{\text{pre}} &= \text{corr}(Y, Y_{\text{pred}} \mid M_X^{\text{pre}}) \\
    \rho_{\text{post}} &= \text{corr}(Y, Y_{\text{pred}} \mid M_X^{\text{post}}) \\
    \Delta \rho &= \rho_{\text{post}} - \rho_{\text{pre}}
\end{align}
The difference $\Delta \rho$ quantifies the intervention-induced change in predictability, providing a natural measure of causal effect. To ensure robust structural identification, we employ a sparsity-constrained DBN learning procedure combining $\ell_1$-regularized and score-based approaches. Proximal gradient optimization identifies salient temporal dependencies while avoiding overfitting. Our framework integrates structure learning, parameter estimation, and causal inference through CCM, enabling quantification of intervention effects in complex temporal systems while maintaining computational tractability and consistent uncertainty treatment across all analysis stages.

By explicitly bridging the geometric and probabilistic dimensions of causal inference, our approach provides a robust framework for identifying causal effects in complex temporal systems. In the next section, we demonstrate the effectiveness of our proposed method, DBN-Informed CCM, through evaluations on real-world datasets.

\section{Experimental Evaluation on Physiological Data}

We evaluated the proposed methodology on pediatric data from 15 children with dystonia (10 acquired, 5 idiopathic/genetic; 9 female) and 13 neurotypical controls (7 female), ages 12–18. Participants performed a grip-maintenance task while receiving mechanical perturbations from an electromechanical tapper per \cite{mcclelland2020abnormal}. Trials were 5 s long with perturbation onset at 1.1 s. Data were preprocessed to remove artifacts and transformed into time-lagged and frequency-domain features.

The framework was implemented in PyTorch 2.1.2 with CUDA 11.2 on an Intel Core i5 (16 GB RAM, macOS). DBN structure learning used an $\ell_1$-regularized, score-based objective optimized via proximal-gradient descent to capture sparse temporal dependencies. Delay embedding reconstructed the shadow manifold to recover state-space dynamics. Kernel-based neighborhood weighting captured nonlinear, distance-decay interactions in state space, while DBN-derived probabilities encoded longer-range temporal influences. These weights were used for cross-prediction to evaluate directional information flow between time series.

\begin{figure}[t]
\centering
\includegraphics[width=0.49\linewidth]{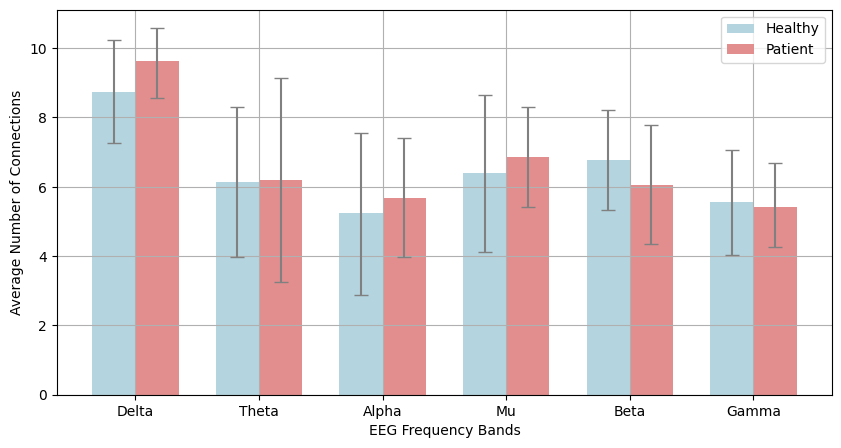}
\includegraphics[width=0.49\linewidth]{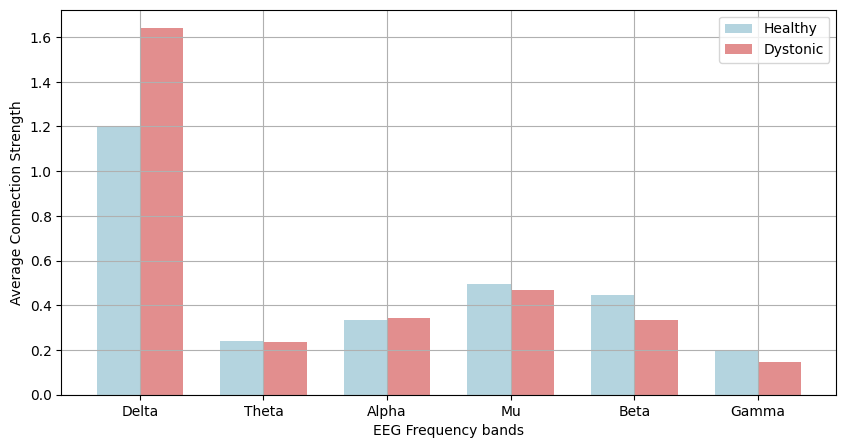}
\caption{Number and strength of connections in healthy and dystonic subjects for different frequency bands.}
\label{conn_strength}
\vspace{-1\baselineskip}
\end{figure}

\begin{figure}[t]
\centering
\includegraphics[width=1.0\linewidth]{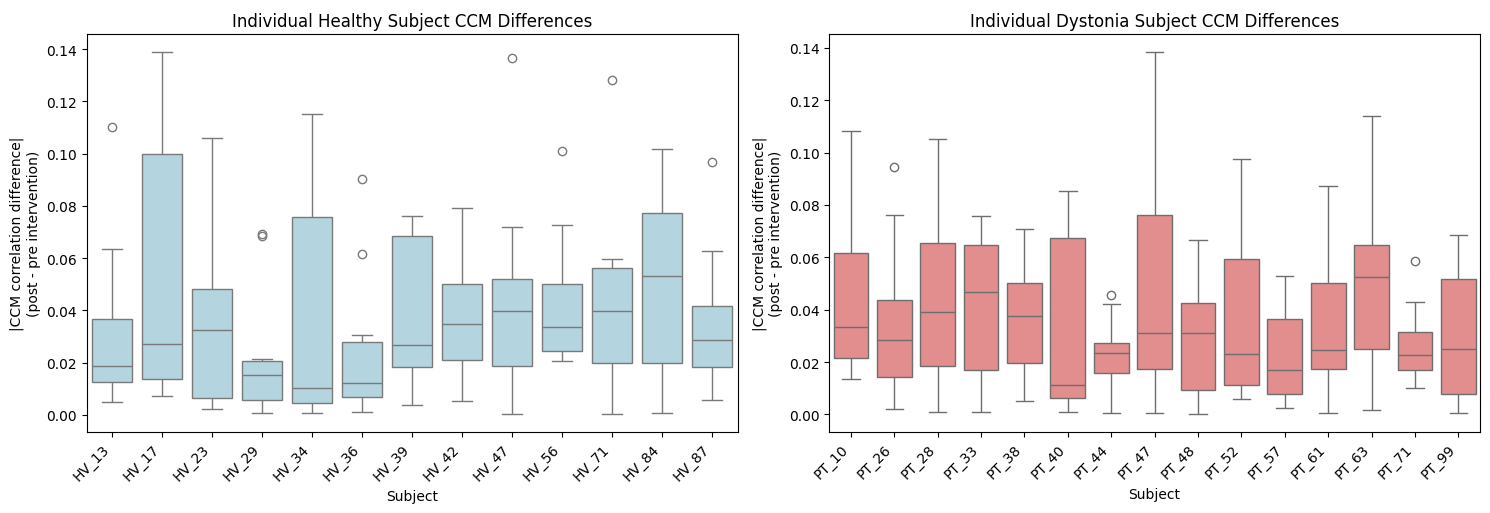}
\caption{Pre and Post Intervention CCM values for all subjects}
\label{all_subs}
\vspace{-1\baselineskip}
\end{figure}

We observe frequency-specific reorganization of connectivity in both groups, reflected in learned DBN connection counts and strengths (Figure \ref{conn_strength}). Dystonic subjects show increased delta-band and decreased beta-band connectivity relative to controls, consistent with impaired motor set and altered cortico-muscular integration; alpha, mu, and gamma differences were smaller and more variable. The error bars across all frequency bands suggest considerable individual variability, necessitating careful interpretation of group-level differences. DBN-informed CCM reveals larger post-perturbation CCM correlation shifts in dystonic subjects (Figure \ref{all_subs}), indicating greater sensitivity and reduced stability of sensorimotor dynamics compared to healthier networks.
Since physiological time series lack ground-truth causal structure, we quantify causal performance using Predictive Consistency (PC) and Causal Impact (CI). PC is the normalized cross-map correlation accounting for shuffled baselines:
\begin{equation}
\text{PC}_{\text{norm}, X \rightarrow Y} =
\frac{ \rho_{\text{pre}}^{X \rightarrow Y} - \rho_{\text{shuffled}}^{X \rightarrow Y} }
{ 1 - \rho_{\text{shuffled}}^{X \rightarrow Y} }
\end{equation}
CI ranks predictors by predictive strength and change under intervention for all nodes in $\setX$:
\begin{equation}
\text{CI}_{X_i \rightarrow Y} =
\rho_{\text{pre}}^{X_i \rightarrow Y} \cdot
\frac{ \big| \rho_{\text{post}}^{X_i \rightarrow Y} - \rho_{\text{pre}}^{X_i \rightarrow Y} \big| }
{ \displaystyle \max_{X_i \in \mathcal{X}} \big| \rho_{\text{post}}^{X_i \rightarrow Y} - \rho_{\text{pre}}^{X_i \rightarrow Y} \big| }
\end{equation}

We computed these metrics across all subject pairs and frequency bands (delta, theta, alpha, beta, mu). Averaged results are reported in Table~\ref{tab:comparison}. The proposed DBN-informed CCM consistently yields higher predictive consistency and interventional stability than linear GC and nonlinear CCM baselines, demonstrating the benefit of combining probabilistic structure learning with manifold dynamics.
\begin{table}[htbp]
\centering
\caption{Comparison of causal inference methods using PCC and ICS metrics (mean $\pm$ std).}
\label{tab:comparison}
\begin{tabular}{lcc}
\toprule
\textbf{Method} & \textbf{PC} & \textbf{CI} \\
\midrule
Granger Causality & $0.62 \pm 0.05$ & $0.68 \pm 0.04$ \\
Standard CCM & $0.63 \pm 0.07$ & $0.71 \pm 0.06$ \\
Gaussian Process CCM  \cite{butler2023causal} & $0.70 \pm 0.05$ & $0.76 \pm 0.04$ \\
Causalized CCM \cite{deng2024causalized} & $0.79 \pm 0.06$ & $0.78 \pm 0.05$ \\
\textbf{DBN-informed CCM (proposed)} & $\mathbf{0.80 \pm 0.03}$ & $\mathbf{0.86 \pm 0.02}$ \\
\bottomrule
\end{tabular}
\vspace{-1\baselineskip}
\end{table}


\section{Conclusion}
This work introduced a causality-aware multimodal framework that bridges probabilistic temporal modeling with nonlinear manifold dynamics to achieve interpretable and robust causal inference in physiological systems. By integrating DBN-based uncertainty estimation with CCM-based geometric analysis, the method effectively captures multiscale brain–muscle interactions. Application to pediatric EEG–EMG data revealed that dystonia reflects frequency-specific network reorganization, marked by enhanced delta and reduced beta coupling, consistent with impaired motor control. Improvements in predictive consistency and causal stability demonstrate the benefit of uniting probabilistic and geometric reasoning. Beyond dystonia, the proposed framework offers a generalizable foundation for multimodal causal analysis in biomedical imaging, with promising applications in real-time neural interfacing, adaptive stimulation, and translational neuroengineering.

\bibliographystyle{IEEEtran} 
\bibliography{References}

\end{document}